# Thermoelectric power of bulk black-phosphorus

E. Flores[1], J.R. Ares[1,*], A. Castellanos-Gomez[2,*], M. Barawi[1], I.J. Ferrer[1], C. Sánchez[1]

The potential of bulk black-phosphorus, a layered semiconducting material with a direct band gap of ~0.3 eV, for thermoelectric applications has been experimentally studied. The Seebeck Coefficient (*S*) has been measured in the temperature range from 300 K to 385 K, finding a value of $S = +335 \pm 10$ µV/K at room temperature (indicating a naturally occurring p-type conductivity). *S* increases with temperature, as expected for p-type semiconductors, which can be attributed to an increase of the charge carrier density. The electrical resistance drops up to a 40 % while heating in the studied temperature range. As a consequence, the power factor at 385 K is 2.7 times higher than that at room temperature. This work indicates the prospective use of black-phosphorus in thermoelectric applications such as thermal energy scavenging, which typically require devices with high performance at temperatures near room temperature.

---

[1] Materials of Interest in Renewable Energies Group (MIRE Group), Dpto. de Física de Materiales, Universidad Autónoma de Madrid, 28049 Madrid, Spain
[2] Instituto Madrileño de Estudios Avanzados en Nanociencia (IMDEA-Nanociencia), 28049 Madrid, Spain
Email: joser.ares@uam.es , andres.castellanos@imdea.org

The recent isolation of few-layer black-phosphorus by mechanical exfoliation of bulk crystals has revived interest from the scientific community in this elemental layered material [1–3], widely studied in its bulk form [4–6]. Few-layer black-phosphorus has already demonstrated potential in nanoelectronic devices and optoelectronic applications. For instance, field-effect transistors made of few-layer black-phosphorus show high performance at room temperature with mobilities of up to 1000 cm$^2$/Vs and current on/off ratios up to 10$^5$ [7–10]. The narrow, direct bandgap of black-phosphorus also makes it interesting for applications requiring a photosensitive material in the near-infrared part of the spectrum [11–16].

Numerous theoretical works, reported shortly after the first experiments on few-layer black-phosphorus devices, predict that black-phosphorus also has great potential in thermoelectric applications [17–21]. According to these works, few-layer black-phosphorus can reach Seebeck coefficients up to 2000 µV/K (at the low doping level). The figure of merit in thermoelectrics *ZT*, defined as $ZT = \sigma \times S^2 T/k$ (with $\sigma$ and $k$ the electrical and thermal conductivity), is theoretically predicted to be in the order of ~1-4 for black-phosphorus which would place this material amongst the best thermoelectric materials to date [18–21]. However, little is experimentally known about the thermoelectric properties of bulk and few-layer black-phosphorus and it seems to be a current subject of debate. In the early works of Bridgman [22] and Warschauer [23] a value for the Seebeck coefficient of bulk black-phosphorus of +413 µV/K and +330 µV/K respectively were quoted, but the obtained data were not shown. Recently, Hong et al. [14] estimated a Seebeck coefficient of +5-100 µV/K (depending on the doping level) from the measurement of photothermoelectric current at 77 K in few-layer black phosphorus samples. Low and co-workers [24] calculated a Seebeck coefficient of ~ +50 µV/K from transconductance measurements of a 100 nm thick black-phosphorus field-effect transistor in combination with effective mass framework calculations of the carrier density [25].

In this work, we present direct measurements of the thermoelectric power of bulk black-phosphorus in a temperature range from 300 K up to 385 K. We find that its Seebeck coefficient at room temperature is +335 ± 10 µV/K, indicating a natural p-type doping (by lattice point defects or uncontrolled impurities) of the crystal. Upon heating, the Seebeck coefficient value increases up to +415 µV/K at 385 K, which could be attributed to an increase of the hole mobility at high temperatures. We also observe an electrical resistance decrease of about 40 % suggesting that the semiconductor is reaching an exhaustion zone and close to the transition from an extrinsic to an intrinsic regimen in the investigated temperature range. At 385 K, the power factor of bulk black-phosphorus is 2.7 times higher than at room temperature. This attribute is highly desirable for thermal waste energy harvesting applications.

Bulk black-phosphorus samples (99.998 % purity) were purchased from Smart Elements, the major supplier of black-phosphorus employed in recent works on this material by different experimental groups [8,9,11,12,26–31]. A bulk sample of black-phosphorus (dimensions approximately 3 mm × 2 mm × 1 mm) has been employed for all the measurements presented here (see Figure 1a), . The bulk sample has been characterized by X-ray diffraction, to determine its crystallinity and orientation, and by infrared spectroscopy. Further characterization of microscopic flakes, cleaved from bulk black-phosphorus crystals, by Raman spectroscopy, optical microscopy, photoluminescence and high resolution transmission electron microscopy have been reported in recent works (see Refs. [9,29,32]). Figure 1b shows an X-ray diffraction pattern from the bulk black-phosphorus sample. The strong intensity of the (020), (040) and (060) diffraction peaks (and almost absence of other diffraction peaks) indicates that the black-phosphorus sample has been mounted with the layers perpendicular to the incident rays. From the X-ray diffraction pattern, one can determine the crystallite size with the Scherrer formula.[33] According to this formula the size of the single-crystalline domains is inversely proportional to the Full Width at Half Maximum (FWHM) of the diffraction pattern peak, after subtracting the instrumental line broadening. From the FWHM of the diffraction peaks of the black-phosphorus sample (barely wider than the instrument resolution), the calculated crystallite size yield values less than 500 nm. From the XRD pattern in Figure 1(b), however, it is not possible to determine the in-plane orientation of the single-crystalline domains as it only gives information about the interplanar distances between [0 h 0] planes. Nonetheless, although several electrical and optical properties of black-phosphorus have proven to be highly anisotropic in-plane, recent density functional theory calculations have predicted a small variation (of less than 1 %) of the Seebeck coefficient value along the different in-plane crystal orientations [19,20,34]).

In order to characterize the thermoelectric properties of bulk black-phosphorus, the sample, shown in Figure 1a, is mounted on a glass substrate and a pair of indium contacts are made to electrically connect the sample. The glass substrate is then placed on an experimental setup designed to measure the thermopower of bulk materials and thin films (see a sketch in Figure 2a)[35]. A pair of aluminum blocks, with embedded heaters, act as thermal reservoirs and allow for varying the average temperature ($T$) and the thermal difference along the sample ($\Delta T$). $\Delta T$ is measured by two thermocouples attached to the indium contacts, placed as close as possible to the black-phosphorus sample. The voltage ($\Delta V$) across the sample is monitored with two voltage probes, also attached to the indium contacts.

Figure 2b shows the thermoelectric voltage generated between the extreme sides of the sample when a temperature difference between them is imposed. The measured voltage scales linearly with the temperature difference showing an almost negligible hysteresis between the temperature ramps. From this measurement, the Seebeck coefficient can be determined as the slope of the linear $\Delta V$ *vs*. $\Delta T$ plot:

$S = \Delta V/\Delta T$        [1]

employing the Telkes criterion [36] for the voltage and the sign of Seebeck coefficient, *i.e.* the Seebeck coefficient has the same sign as the charge carriers: positive for p-type and negative for n-type samples. The room temperature value determined for the bulk black-phosphorus sample is $S = +335 \pm 10$ µV/K. The positive sign of the Seebeck coefficient indicates that the bulk black-phosphorus sample is a naturally occurring p-type semiconductor, in agreement with the observed behavior of multilayered black-phosphorus flakes.[7–9,11,14,28,37,38] Moreover, the measured value agrees with that quoted by Bridgman [22] and Warschauer [23], also measured in bulk samples. Interestingly, our result also agrees with that calculated by Lv *et al*. [34] for bulk black-phosphorus crystals assuming a hole doping level of $10^{12}$ cm$^{-2}$, which is a reasonable doping level.[39] The values indirectly determined from transconductance and photocurrent measurements in exfoliated black-phosphorus samples are 3 to 60 times smaller than those determined for bulk samples or calculated for bulk and single-layer black phosphorus. Therefore, a direct measurement of the Seebeck coefficient of thin exfoliated flakes would be necessary to determine whether this discrepancy is due to an effect of the reduced dimensionality (stronger influence of the substrate or environment due to larger surface-to-volume ratio) of exfoliated flakes or if it is due to uncertainties involved in the estimation process. Table 1 summarizes the results of different experimental works on the thermoelectric properties of black-phosphorus.

| Sample | Seebeck coeff. (µV/k) | Temperature (K) | Reference |
|---|---|---|---|
| Bulk | +335 to +415 | 300 to 385 | This work |
| Bulk | +413 | 300 | Bridgman *et al*. [22] |
| Bulk | +330 | 300 | Warschauer *et al*. [23] |
| 8 nm thick | +5 to +100 (*) | 77 | Hong *et al*. [14] |
| 100 nm thick | +50 (*) | 300 | Low *et al*. [24] |

Table 1: Comparison between the Seebeck coefficient obtained in this work with other reported experimental values. Values marked with (*) have been indirectly determined.

To further characterize the thermoelectric properties of black-phosphorus, $S$ has been measured as a function of $T$ up to 385 K. The base temperature is controlled by heating both aluminum blocks at once. Figure 3a shows the Seebeck coefficient determined at different average temperatures in the range of 300 K to 385 K. $S$ monotonically increases in the studied temperature range. It seems to saturate around a value of +415 µV/K, suggesting that an exhaustion zone has been reached. In fact, the transition between the extrinsic and intrinsic regime in black phosphorus occurs around 380 K [4].

In a semiconductor the maximum value of the Seebeck coefficient is reached at the transition between the extrinsic and intrinsic regime which is closely related to the energy gap ($E_g$) and to the temperature of the transition ($T$) by the following equation

$S_{max} = E_g/2qT$        [2]

with $q$ is the electron charge [40]. Using the maximum value of the Seebeck coefficient (+415 µV/K) at $T$ = 380 K in Equation 2, a $E_g$ for black phosphorous of 0.32 eV is obtained. This value is similar to that reported by Asahina *et al*. [41] and it has been confirmed for the samples used in this work by Infrared Spectroscopy measurements (not shown here).

Within the studied temperature range we have also measured the change in the electrical resistance to gain deeper insight on the temperature dependence of the thermoelectric performance of black-phosphorus. Note that due to the irregular geometry of the studied sample it is difficult to accurately determine the electrical resistivity. Figure 3b shows the measured electrical resistance at different temperatures in the range of 300 K to 380 K. The resistance decreases when the temperature increases, as expected for a semiconducting material. The decrease in resistance reaches 40 % at 380 K. The combination of increased Seebeck coefficient and decreased electrical resistance observed at higher temperature means that the power factor of black-phosphorus is enhanced at higher temperatures. The thermoelectric power factor is defined as:

$$PF = S^2/\rho = S^2/A \times L \times R \qquad [3]$$

where $\rho$ is the electrical resistivity, $A$ is the cross sectional area perpendicular to the electrical transport direction and $L$ is the channel length. Figure 3c shows the temperature dependence of the power factor, normalized to the room temperature value. As can be observed, the power factor increases with temperature up to a value 2.7 times higher than that at room temperature. This fact is not only due to the increasing dependence of the Seebeck coefficient on temperature, which dominates the power factor, but also to the decrease of the resistivity as temperature is increased. This result confirms the suitability of black-phosphorus as a thermoelectric material at moderately high temperatures.

In summary, we have experimentally characterized the performance of black-phosphorus for thermoelectric applications. The Seebeck coefficient has been directly determined in the temperature range of 300 K to 385 K. At room temperature, the Seebeck coefficient is $S = +335 \pm 10$ µV/K, indicating a naturally occurring p-type doping in bulk black-phosphorus samples. $S$ increases with temperature reaching a saturation value at $T = 380$ K indicating that a transition between the extrinsic and intrinsic regime occurs. The electrical resistance has been measured at different temperatures in order to determine the temperature dependence of the thermoelectric power factor. At moderately high temperatures, the power factor increases to 2.7 times the room temperature value. This work explores the potential of black-phosphorus in thermoelectric applications, especially in those requiring temperatures above room temperature such as thermal waste energy harvesting.

ACKNOWLEDGMENTS

The authors thank Joshua O. Island (Delft University of Technology) for carefully reading the manuscript and insightful discussions. Authors from MIRE Group acknowledge the support of the Ministry of Economy and Competitiveness (MINECO) for this research (contract MAT2011-22780). They also thank technical support from Mr. F. Moreno. E. Flores acknowledges to the Mexican National Council for Science and Technology (CONACyT) for providing the funding necessary to complete his PhD. A.C-G. acknowledges financial support through the FP7-Marie Curie Project PIEF-GA-2011-300802 ('STRENGTHNANO') and the Fundación BBVA through the grant 'I Convocatoria de Ayudas Funcación BBVA a Investigadores, Innovadores y Creadores Culturales' ('Semiconductores ultradelgados: hacia la optoelectrónica flexible').

FIGURES

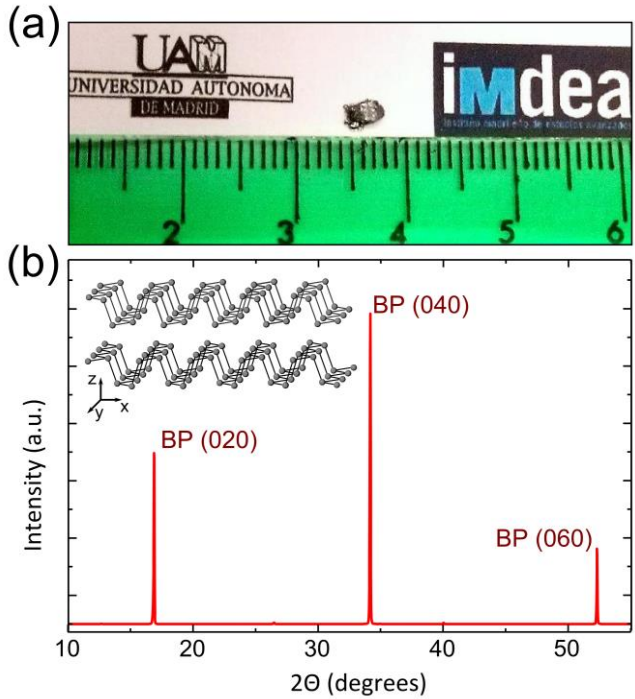

FIG. 1. (a) Optical image of the black-phosphorus bulk crystal employed in this study. (b) X-ray diffraction spectrum of the black-phosphorus sample. The X-ray diffraction data has been also used to mount the sample with the phosphorus layers parallel to the sample holder, as evidenced by the strong intensity of the (020), (040) and (060) peaks and the lack of the other diffraction peaks in the X-ray diffraction pattern.

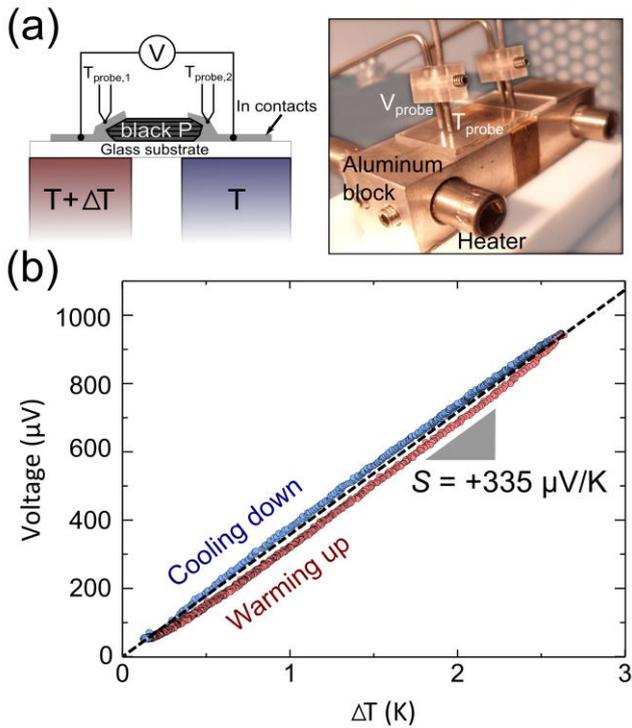

FIG. 2. (a) Sketch and picture of the experimental setup employed to determine the Seebeck coefficient of bulk black-phosphorus. (b) Thermoelectric generated voltage ($\Delta V$) as a function of the temperature difference along the black-phosphorus sample during a warming up and cooling down cycle. The Seebeck coefficient can be extracted from the slope $S = + 335 \pm 10$ µV/K.

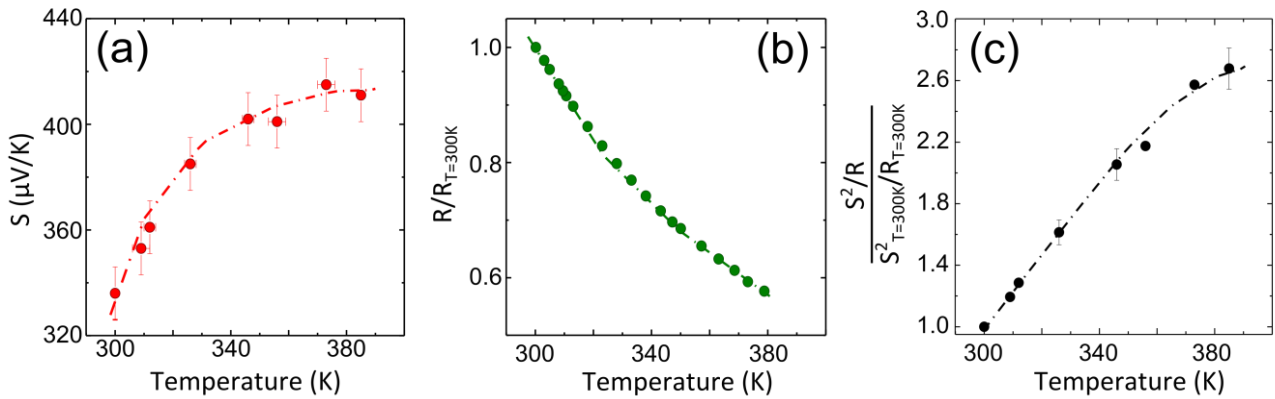

FIG. 3. Temperature dependence of the Seebeck coefficient (a), electrical resistance (normalized to the room temperature value) (b) and power factor (normalized to the room temperature value) (c) of bulk black-phosphorus.

13